\documentclass{openjournal}

\usepackage{lipsum}

\usepackage{xcolor}
\usepackage{textgreek}
\usepackage[utf8]{inputenc}
\usepackage[english]{babel}

\usepackage{hyperref}
\hypersetup{
    unicode, 
    colorlinks=true,
    linkcolor=linkcolor,
    citecolor=linkcolor,
    filecolor=linkcolor,
    urlcolor=linkcolor,
}
\usepackage{color,colortbl}
\definecolor{linkcolor}{rgb}{0.0,0.3,0.5}
\usepackage{tensind}
\tensordelimiter{?}
\DeclareGraphicsExtensions{.bmp,.png,.jpg,.pdf}
\usepackage{verbatim}

\usepackage{orcidlink}
\usepackage{soul}

\usepackage{graphicx}%
\usepackage{multirow}%
\usepackage{amsmath,amssymb,amsfonts}%
\usepackage{amsthm}%
\usepackage{mathrsfs}%
\usepackage[title]{appendix}%
\usepackage{xcolor}%
\usepackage{textcomp}%

\usepackage{booktabs}%
\usepackage{algorithm}%
\usepackage{algorithmicx}%
\usepackage{algpseudocode}%

\graphicspath{ {./figs/} }

\begin{document}

\title[Tidally Delayed Spin-Down of Very Low Mass Stars]{ Tidally Delayed Spin-Down of Very Low Mass Stars}





\author{Ketevan Kotorashvili\orcidlink{0000-0002-5706-1299}}
\email{kkotoras@ur.rochester.edu}
\affiliation{Department of Physics and Astronomy, University of Rochester, 500 Wilson Boulevard, Rochester, 14627, NY, USA.}
\affiliation{Laboratory for Laser Energetics, University of Rochester, 250 East River Road, Rochester, 14623, NY, USA. }

\author{Eric G. Blackman\orcidlink{0000-0002-9405-8435}}
\email{blackman@pas.rochester.edu}
\affiliation{Department of Physics and Astronomy, University of Rochester, 500 Wilson Boulevard, Rochester, 14627, NY, USA.}
\affiliation{Laboratory for Laser Energetics, University of Rochester, 250 East River Road, Rochester, 14623, NY, USA. }

\begin{abstract}

Very low-mass  main-sequence stars reveal some curious trends in  observed rotation period distributions that require abating the spin-down  
 that standard rotational evolution models would otherwise imply. 
By dynamically
coupling magnetically mediated spin-down to tidally induced spin-up from close orbiting substellar companions, 
we show that tides from sub-stellar companions may explain these trends.
In particular, brown dwarf companions 
can delay  the spin-down
and explain the dearth of field, late-type M dwarfs with intermediate rotation periods. 
 We find that tidal forces also strongly influence stellar X-ray activity evolution, so that methods of gyrochronological aging must be generalized for stars with even sub-stellar companions. We also discuss how the theoretical predictions  of the 
 spin evolution model can be used  with future data to constrain the population distribution of companion orbital separations.
\end{abstract}

\begin{keywords}
    {stars: magnetic field, -- stars: rotation, -- stars: late-type -- stars: low-mass,  -- planet–star interactions, -- brown dwarfs,}
\end{keywords}

\maketitle

\section{Introduction}
\label{sec:intro}

The observed period distributions of main sequence stars as a function of age constrain models of stellar spin evolution.  In turn, understanding the spin evolution  is important for gyrochronological aging \citep{barnes2003rotational, Reiners2012}, which depends on a one-to-one mapping between rotation period and stellar age
bolstered by a predictive theory. 
However, observations reveal that the rotation  and  activity of significant populations of main sequence stars evolve anomalously with age 
\citep{meibom2009stellar,meibom2011kepler,agueros2018new}, 
challenging standard theoretical models restricted to the magnetically mediated spin down of isolated stars \citep{Matt2012,Reiners2012,vanSaders2013,Gallet2013improved,Matt2015,vanSaders2016}.
 \citep{Newton2016}. Observations from multiple surveys \citep{irwin2011Angular,Newton2016,Kado-Fong2016,Newton2018,Howard2020,popinchalk2021evaluating,Shan2024carmenes} 
consistently reveal 
a bimodal distribution in rotation period for M dwarfs with a notable
dearth of late-type M dwarfs with intermediate rotation periods.
This bimodality is characterized  by populations of rapid and slow rotators ($>100$ days) with  a significant lack of intermediate rotators between roughly 10-40 days \citep{Kado-Fong2016} and in some cases up to 50 \citep{Shan2024carmenes} or 70 days \citep{Newton2016}. This gap can be interpreted as evidence of a rapid spin-down of M dwarfs. 

While a change in magnetic topology  might help to explain some anomalies \citep{Metcalfe2016, Garraffo2018}, many stars have  stellar, substellar, or planetary companions whose  influence on spin angular momentum evolution  must also be understood. 
  M dwarfs with short rotation periods 
 may either be very young, or involve  short-period stellar binaries \citep{McQuillan2013}.  Using galactic kinematics to estimate M dwarf ages,  there is in fact a sub-population of fast-rotating old stars ($ 7-13 $ Gyr) \citep{irwin2011Angular}.
The presence of a sufficiently massive companion can modify the spin evolution  of the host stars via equilibrium and dynamical tides \citep{Zahn1977,Ogilvie+lin2007,Ogilvie2013,Mathis2015b,benbakoura2019evolution,ahuir2021dynamical,Pezzotti2025}  and we shall see that this can
 generate the gap
in the rotation period distribution. 
Even sub-stellar brown dwarf (BD)  companions may be sufficient to explain this trend for the sub-population of old, late-type M stars.


In contrast to  semi-empirical \citep{Matt2015} or semi-analytical and numerical magnetohydrodynamic 
models \citep{Gallet2013,Gallet+Bouvier2015,Amard2016, Garraffo2018,benbakoura2019evolution,ahuir2021dynamical,Pezzotti2025}, our model herein is derived from  first principles, starting with a dynamical MHD approach \citep{Blackman+owen2016} that is   augmented to include tides. While existing models typically use parametrized torque laws based on MHD simulations or empirically tuned to match observations, they often treat stellar properties as independent or use external prescriptions and empirical scaling laws to couple them. Our model dynamically and self-consistently couples the evolution of X-ray luminosity, rotation, mass loss, magnetic field strength to tidal effects of the companion and its orbital evolution. 
Previous work has also focused on solar-type stars, the effects of core-envelope decoupling and dynamical tides in addition to equilibrium tides in studying the influence of a companion on the stellar surface rotation and X-ray activity \citep{Gallet2013,Gallet+Bouvier2015,Amard2016, Rao2021,benbakoura2019evolution,ahuir2021dynamical,Pezzotti2025}.
Here we study the influence of tides on the stellar spin evolution of fully convective M dwarf-BD systems for a range of companion masses and orbital separations using a simple prescription for tidal torque. We focus on explaining the dearth of intermediate rotators 
for late-type M dwarfs \citep{Newton2016} by accounting for equilibrium tides.
For comparison, we also investigate the spin evolution implications of 
sub-BD mass companions for partially convective K dwarfs and early M dwarfs. 

\section{Methods}\label{sec:methods}



\subsection{Magnetic spin-down model for an isolated magnetized star}\label{subsec:mb}
Our spin evolution model that dynamically couples stellar spin, magnetic field strength, X-ray luminosity and mass loss is
based on the   analytic framework of \citep{Blackman+owen2016}, where the original model was derived and applied to the Sun,  and was further developed  in  \citep{kbo2023}  for application to  older-than-solar low-mass MS stars. In this conceptual framework,  dynamo-generated magnetic fields are the main coronal energy source, supplying both open and closed coronal fields of strength proportional to each other. Hot gas propagates along open field lines and removes angular momentum, while X-ray emission from magnetic dissipation results from reconnecting closed field lines that can subsequently open up.  
The main dynamically coupled  ingredients  are:  an isothermal Parker  wind \citep{Parker1958};
angular momentum loss from the equatorial plane 
\citep{Weber1967};  a dynamo saturation prescription based on a combination of magnetic helicity evolution and  buoyant loss of magnetic fields \citep{Blackman2002, Blackman2003}, used to estimate the magnetic field strength; prescription for the turbulent correlation time  in the dynamo coefficients that asymptotes to the convection time for slow rotators and shear or rotation time for very fast rotators \citep{Blackman2015}; and a  coronal equilibrium condition \citep{Hearn1975} that determines how the  magnetic energy supply is divided into radiation, wind, and conductive losses. Together, these components lead to a system of coupled equations for the X-ray luminosity, magnetic field, spin evolution and mass loss below.  
More detailed  derivations  are in \citep{Blackman+owen2016} and   \citep{kbo2023}. 

The first key equation that emerges from the model is the relation between dimensionless X-ray luminosity $l_{\rm x {\star}}(t)$ and radial magnetic field $b_{r{\star}}(t)$:
\begin{equation}
    l_{\rm x {\star}}(t) \equiv g_L(t)\left(\frac{\rm{erf}(\epsilon/\Tilde{R}o_{\star})}{\rm{erf}(\epsilon/\Tilde{R}o(t))}\right)^{\frac{4}{3(1-\lambda)}}\left(\frac{1+\Tilde{R}o_{\star}/\textrm{erf}(\epsilon/\Tilde{R}o_{\star})}{1+\Tilde{R}o(t)/\textrm{erf}(\epsilon/\Tilde{R}o(t))}\right)^{\frac{2}{1-\lambda}}=b_{r{\star}}(t)^{^{\frac{4}{1-\lambda}}}.
	\label{eq:lxbr}
\end{equation}


This surface radial magnetic field
originates from dynamo-produced toroidal interior field that emerges to the surface.  The  $b_r$
is interpreted as the azimuthally and dynamo cycle-averaged poloidal field which we scale as local  radial field.The associated X-ray luminosity is a product of the magnetic energy flux averaged over a dynamo cycle times the solid angle fraction  $\Theta/4\pi$ through which magnetic field rises, which approximately equals the areal fraction of sunspots during the cycle.
Here 
$ \lambda$ is the exponent of  the power law dependence of  the magnetic starspot area covering fraction $\Theta$  on X-ray luminosity $\mathcal{L}_X$, namely
 $\Theta\propto \mathcal{L}_X^\lambda$. We  take $\lambda=1/3$, which is consistent with the range inferred from observations of star spot covering fractions for FGKM stars \citep{NicholsFleming2020}. 
The field that emerges over a surface area consistent with the sunspot covering fraction $\Theta$ is $|B_{r\star }|=\Theta b_{r\star }\simeq \mathcal{L}_X^\lambda b_{r\star }\propto l_{\rm x {\star}}^\lambda b_{r\star }\propto b_{r{\star}}^{(\frac{4}{1-\lambda})\lambda}b_{r\star }$. For $\lambda=1/3$ we have $|B_{r\star }|\propto b_{r{\star}}^{3}$ and $l_{\rm x {\star}}\propto|B_{r\star }|^2.$, which is in a good agreement with observed relationship between magnetic flux and X-ray luminosity \citep{Pevtsov2003,Vidotto2014,See2015,Zhuleku2020}.

The generalized turbulent correlation time scale for  turbulence of stellar dynamo theory comes from the physical argument that  for slow rotators, the convection time determines eddy correlation but for fast rotators eddies can be radially sheared  faster than convection can overturn them.  The results are sensitive to the distinction between the two asymptotic regimes and less so to the form of the transition between them. Compared to the original model of \citep{Blackman+owen2016}, we have modified the transition formula:
instead of using a shear parameter\footnote{ $|\Omega_0-\Omega(r_c,\theta_s)|=\Omega_0/s$, where $\Omega$ is surface rotational speed; $\theta_s$ is a fiducial polar angle; $r_c$ is a fiducial radius in the convective zone.} $s$
we simply use the error function  $\rm erf(\epsilon/\Tilde{R}o_\star)$  which depends on the parameter
$\epsilon$ and Rossby number $\Tilde{R}o_\star$. We  estimate $\epsilon$ 
$0.009\leq\epsilon\leq 0.0036$ 
which is consistent with model-inferred Rossby numbers that account for the the observed transition from the saturated to the unsaturated regime of X-ray luminosity
of both partially and fully convective stars \citep{Wright2011,Reiners2014,Wright2018} and the range accounts for the corresponding uncertainty in the saturation Rossby number. To estimate the Rossby number we use the convective turnover timescale ($\rm \tau_c$) prescription given in \citep{Wright2018}. All the fixed model parameter definitions, values and prescriptions are summarized below in Table \ref{tabparameters}.
Second, we use a dynamo saturation model, which gives an estimate for the saturated large scale magnetic field strength based on the combination of magnetic helicity evolution and loss of magnetic field by magnetic buoyancy \citep{Blackman+2000, Blackman2002}. Here $g_L(t)=({1.4-0.4t})^{-\frac{\lambda-1}{4}}$  captures the bolometric luminosity evolution on the main sequence. 
For simplicity, we take $g_L(t) =1$ for both partially and fully convective stars. 

Next,  we consider angular momentum loss from equatorial plane following \citep{Weber1967,Lamers1999} to find the azimuthal component of magnetic field at the base of the corona which we refer to as the "coronal toroidal field" and the equation for angular velocity. We assume a  Parker spherical wind  \citep{Parker1958} which propagates along radial large-scale fields through   the Alfv\'en radius $r_A(t)$, which following \citep{Weber1967,Blackman+owen2016,Lamers1999} is given by, 
\begin{equation}
\frac{r_A(t)}{R_\star }=\left(1-\frac{R_\star B_{r,\star i}B_{\phi,\star i}}{\dot{M_{\star i}}\Omega_{\star i}}\right)^{1/2}=\left(1+\frac{R_\star |B_{r,\star i}||B_{\phi,\star i}|}{\dot{M_{\star i}}\Omega_{\star i}}\right)^{1/2},
\label{eq:ra}
\end{equation}
where quantities $ \dot{M}_{\star i}$, $B_{\rm r,\star i}$, $B_{\rm \phi,\star i}$, $\Omega_{\star i}$  are the initial mass loss,   radial magnetic field at the coronal base,  coronal toroidal magnetic field, and angular velocity, respectively. Here  $R_{\star }$ is the stellar radius and assumed to be constant since we are focusing on stars already on the main sequence.

Combining equation (\ref{eq:ra}) with a separate expression for $r_A(t)$  that is independent of coronal toroidal magnetic field, we obtain the coronal toroidal magnetic field equation.   As derived from the mass loss rate to outflow speed relation,  the definition of the Alfv\'en radius, and the radial field fall off of $1/r^2$, $r_A(t)$ is given by
\begin{equation}
\frac{r_A(t)}{R_{\star}}=\frac{b_{r{\star}}(t)}{\Dot{m}_\star(t)^{1/2}\Tilde{u}_A(t)^{1/2}}\frac{R_{\star}B_{r,{\star}i}}{\Dot{M}^{1/2}_{{\star}i} u^{1/2}_{A,{\star}i}},
\label{eq:ra2}
\end{equation}
where $\dot{m}_\star(t)$ is the mass loss rate (see equations \ref{eq:lxmdot1} and  \ref{eq:lxmdot2} below); 
$\Tilde{u}_A(t)$ and $u_{A,{\star}i}$ are the dimensionless Alfv\'en speed and initial value for Alfv\'en speed given in equation (\ref{eq:ua}) below. 
Combining equations (\ref{eq:ra}) and (\ref{eq:ra2}) gives
\begin{equation}
 b_{\phi {\star}}(t)\equiv\frac{B_{\phi{\star}}(t)}{B_{\phi ,{\star}i}}=-\frac{\dot{m}_\star(t)\omega_{\star}(t)}{b_{r_{\star}}(t)} \frac{M_{{\star}}\Omega_{{\star}i}}{R_{\star}B_{\phi,{\star}i}B_{r,{\star}i}}\left[\frac{r_A(t)^2}{R^2_{\star}}-1\right],
	\label{eq:bfi}
\end{equation}
where $b_{\phi {\star}}(t)$ is the dimensionless coronal toroidal magnetic field, normalized to the initial coronal toroidal magnetic field $B_{\phi,\star i}$
\footnote{We substitute subscript $\rm i$ for  $\rm n$ used in \citep{Blackman+owen2016}, as here we use initial stellar values for rather than present-day values.}. 
$M_{\star }$ is the stellar mass and assumed to be constant; 
$\omega_{\star }(t)=\Omega_\star(t)/\Omega_{\star i}$ is the dimensionless angular velocity. From conservation of angular momentum in the \cite{Weber1967} framework, for which  the magnetic field is nearly rigid out to the Alfv\'en radius,  the stellar angular velocity evolves as 
\begin{equation}
    \frac{d\Omega_\star(t)}{dt} = -\frac{q \dot{M}_{\star}(t)\Omega_{\star }(t)}{\beta^2 M_{\star}} \frac{r_A(t)}{r_\star}^2
\end{equation}
and dimensionless angular velocity is described by
\begin{equation}
    \frac{d{\omega}_{\star}(t)}{d\tau} = - {\omega}_{\star}(t) t_{{\star}}\frac{q}{\beta_\star^2}\frac{b_{r \star}(t)^2}{ m_{\star} \Tilde{u}_A(t)}\frac{B^2_{r,{\star}i}R^2_{\star}}{M_{{\star}}u_{A,{\star}i}},
	\label{eq:omegadot wind}
\end{equation}
where $t_\star$ is the normalization parameter taken as $1$ Gyr. $m_{\star}=1$ is the dimensionless stellar mass in units of initial stellar mass, which does not evolve. Here $q$ is a dimensionless moment of inertia parameter that depends on the  assumed constant fraction of the star in which the field is anchored. We use $q=1$ for all stars, which is equivalent to assuming that  the full stellar mass is coupled to the anchoring magnetic field that contributes to  spin-down; $\beta_\star$ is the radius of gyration taken from stellar evolutionary models computed using the Modules for Experiments in Stellar Astrophysics package (MESA) \citep{Claret2023}. For a stellar mass $M_\star=0.2M_\odot$ we use $\beta_\star= 0.458$, while for $M_\star=0.6M_\odot$ we take $\beta_\star= 0.359$.

Time dependent data are needed to accurately assess the consistency of equation (\ref{eq:bfi}) but we can get an approximate time averaged prediction by using our numerical solutions to find the scaling between toroidal and observed radial magnetic fields $b_\phi(t)=Cb_r(t)^{\zeta(t)}\propto C|B_r(t)|^{\zeta(t)/3}$
where the proportionality follows from the discussion below equation (\ref{eq:lxbr}), and we infer $\zeta(t)$  by solving for $b_\phi(t)$   as a function of $b_r(t)$. If we approximate $\zeta(t)$ from $t\approx
 1.5-10$ Gyr 
 by a constant for the systems considered in our paper,  we find that $C$ - $[10^{-1}, 10^{-3}]$ and $\eta=\zeta/3$ falls into the range between $[0.7, 1]$,  roughly comparable to the observed scalings between toroidal field energy and poloidal field energy.  \citep{See2015}. More detail on our estimate for $\eta$ is given in the supplement Fig. \ref{fig:suplfit}. 
 
 In this model,  dynamo generated magnetic fields source the energy for the wind and X-rays or are dissipated through thermal conduction. To determine the dominant flow path of magnetic energy we use the  coronal flux equilibrium condition \citep{Hearn1975} between wind flux, conductive loss and the radiative (X-ray) loss, assuming that the equilibrium is established over shorter time-scales compared to the secular gigayear evolution time. For detailed derivations refer to  \cite{Blackman+owen2016} and \cite{kbo2023} where we find the relation between mass loss and X-ray for two regimes: regime I as the lifetime phase of a star for which thermal conduction flux dominates the outflow flux and
regime II when the reverse is true.
For the thermal conduction loss dominated regime - Regime I:
\begin{equation}
\Dot{m}_\star(t) \simeq l_{\rm x \star}(t)^{\frac{23}{38}} \exp \bigg[\frac{3.9}{\Tilde{T}_{0{\star}}}\frac{m_{\odot \star}}{r_{\odot \star}}\bigg(1-(l_{\rm x \star}(t)^{-\frac{16}{9}}\bigg)\bigg],
\label{eq:lxmdot1}
\end{equation}
whilst for the  wind loss dominated regime - Regime II:
\begin{equation}
   \Dot{m}_\star(t)\simeq l_{\rm x \star}(t) \simeq \exp \bigg[\ln(\Tilde{T}_{0{\star}}) +\frac{7.8}{\Tilde{T}_{0{\star}}}\frac{m_{\odot \star}}{r_{\odot \star}}\bigg(\frac{\Tilde{T}_{0{\star}}}{\Tilde{T}_{0,{\star}i}}-1\bigg) \bigg], 
	\label{eq:lxmdot2}
\end{equation}
where $\Tilde{T}_{0\star }(t)$ is a dimensionless coronal temperature at equilibrium, normalized such that transition between regime I and regime II occurs at $\Tilde{T}_{0\star }=0.5$. The $\Tilde{T}_{0,{\star}i}$ represents the initial coronal temperature.
\footnote{Here we have included the correction \citep{kbo2023errata} to  \citep{kbo2023}.} 
Here 
$m_{\odot \star}=\frac{M_{\star}}{M_{\odot}}$ and  $r_{\odot \star}=\frac{R_0}{R_{\odot}}\sim\frac{R_\star}{R_{\odot}}$, where $R_0$ represents radius at the coronal base; $M_{\odot}$ and $R_{\odot}$ are solar mass and radius respectively.  \cite{Blackman+owen2016} studied the wind dominated regime for the Sun, and found that equation (\ref{eq:lxmdot2}) is consistent with the data.

Alfv\'en speed is given by
\begin{equation}
    \Tilde{u}_A(t) \equiv \frac{u_A(t)}{u_{A,{\star}i}}= \sqrt{\frac{T_{0\star}(t)}{T_{{0, \star}b}}}
    \frac{W_h[-D(r_A(t))]}{W_h[-D(r_{A,{\star}i})]},
	\label{eq:ua}
\end{equation}
where $W_h[-D(r_A)]$ is the Lambert W function for the Parker wind \citep{Parker1958} with $h= 0$ for $r\leq r_s$ and $h= -1$ for $r\geq r_s$ \citep{Cranmer2004}, and 
\begin{equation}
   D(r_A)=\left(\frac{r_A}{r_s}\right)^{-4} {\rm exp} \left[4\left(1-\frac{r_s}{r_A}\right)-1\right].
\end{equation}
where the sonic radius is given by
\begin{equation}
 \frac{r_s}{R_{\star}}=\frac{GM_\star}{2c^2_sR_{\star}}  
 \label{eq:rs}
\end{equation}
where $G$ is the gravitational constant and $c_s$ is isothermal sound speed.



\subsection{Influence of tides on the evolution of stellar spin 
}\label{sec:tides}

The spin evolution model summarized in the previous section must be generalized to account for tidal stresses, which
can exchange angular momentum  between the  orbit and the spins of the primary and secondary when the  primary star has a close stellar or sub-stellar companion. 
We assume that   tidal stresses  are redistributed much faster than the secular spin evolution that we will solve for 
\citep{Zahn2008,bookrozelot2012environments}.
 We focus only on  equilibrium tides, for which the star   evolves  in hydrostatic equilibrium  during its response to tidal forcing. Dynamical tides can enhance the tidal influence but compared to equilibrium tides they become relevant only for a short time fraction of the overall evolution.
Including these effects can be a subject for future work.
 \citep[e.g.,][]{Zahn1977,tassoul_2000}.




We consider a sub-stellar companion in a circular orbit, 
and  a uniform  rotation of the primary star at angular velocity $\Omega_{\star}$. 
The  gravitational field of the secondary deforms the primary, and vice-versa, but we focus on the former.
Since the star is not a perfect fluid,
it will incur a viscously mediated  
time lag $\Delta t$ to the deformation that  
offsets the direction of
tidal bulges from the 
instantaneous  direction of 
the external force  by a small angle $\delta$,  producing a torque. We adopt the weak friction approximation. In the weak-friction approximation $\delta\ll 1$, assuming that the tidal force acts as a perturbation on the primary. Then the tidal torque $\Gamma_{\rm T}$  on the primary can be written \citep[e.g.,][]{tassoul_2000}:
\begin{equation}
    \Gamma_{\rm T}\simeq 
   - 3k\frac{Gm^{2}}{R_{\star}}\left(\frac{R_{\star}}{a}\right)^6 \delta,
    \label{eq: tidal torque1}
\end{equation}
where 
$m$ is the secondary (companion) mass 
and $a$ is the orbital separation.
Here $k$ is the apsidal motion constant  determined by the structure of the star, where 
we use values $\rm k=0.155$ and $\rm k=0.054$ from MESA \citep{Claret2023} corresponding to the stellar masses $M_\star=0.2M_\odot$ and $M_\star=0.6M_\odot$.


In the weak friction approximation,
\begin{equation}
  \delta=(\Omega_{\star} - \Omega_{\rm orb}) \Delta t=(\Omega_{\star} - \Omega_{\rm orb})\frac{t^2_{\rm ff}}{t_{\rm conv}},
  \label{eq: tidal lag delta}
\end{equation}
showing that
$\delta$ is linearly proportional to  the departure from synchronism and a constant factor $\Delta t=t_{\rm ff}^2/t_{\rm conv}$, the ratio of 
the free-fall time squared to the convective friction timescale.
Here   $t_{\rm ff}=\left({GM_\star}/{R_{\star}^3}\right)^{-1/2}$  with  orbital angular velocity $\Omega_{\rm orb}$  given by: 
\begin{equation}
    \Omega_{\rm orb}=\left[\frac{G(M_{\star}+m)}{a^3}\right]^{1/2}.
    \label{eq: omega orb}
\end{equation}
The time $t_{\rm conv}$ is a characteristic timescale for kinetic energy dissipation  in the stellar convection zone  governed by turbulent viscosity
and can be written as  $t_{\rm conv}= R_{\star}^2/\nu_t\simeq\left(M_{\star}R_{\star}^2/L_{\star}\right)^{1/3}$ 
,  where $\nu_t$ is eddy viscosity and $L_{\star}$ is the stellar  bolometric luminosity 
\citep{Zahn1989}. This is a crude but useful order-of-magnitude scaling for studying tidal dissipation in convective envelope.

Combining equations (\ref{eq: tidal torque1}) and (\ref{eq: tidal lag delta}) gives the 
evolution of stellar angular velocity due to tides: 
\begin{equation}
    \Gamma_{\rm T}= I_\star\left.\frac{d\Omega_{\star}}{dt}\right|_{\rm T}= 
    - 3k \frac{(\Omega_{\star} - \Omega_{\rm orb})}{t_{\rm conv}}\left(\frac{m}{M_{\star}}\right)^2 M_{\star} R_{\star}^2 \left(\frac{R_{\star}}{a}\right)^6,
    \label{eq: tidal torque }
\end{equation}
where $I_\star$ is the moment of inertia of the star. 
Equation (\ref{eq: tidal torque }) shows that only when the stellar rotation is not synchronized with the orbital motion so that $\Omega_{\star} \ne \Omega_{\rm orb}$, does
a finite $t_{\rm conv}$  cause
a lag of the tidal bulges $\delta \neq 0$, and in turn, a tidal  torque.

Considering only  tidal interaction without magnetic braking,  conservation of  total angular momentum 
of the  star and companion gives
 $d\left[I_\star\Omega_\star(t)+ma^2(t)\Omega_{\rm orb}(t)\right]/dt=0$ and correspondingly, 
 the evolution of orbital separation 
\begin{equation}
    \frac{d a(t)}{dt}=-a(t)^{1/2}\Gamma_{\rm T}(t)\frac{2}{m(G(M_\star+m))^{1/2}}
    \label{eq: adot },
\end{equation}
which can be rewritten in dimensionless form as
\begin{equation}
    \frac{d \Tilde{a}(t)}{d\tau}=-\Tilde{a}(t)^{1/2}\gamma_{\rm T}(t)\frac{2\beta_\star^2}{q}\frac{M_\star}{m}\left(\frac{\chi M_\star}{M_\star+m}\right)^{1/2},
    \label{eq: adot_tilda }
\end{equation}
where $   \gamma_{\rm T}= \left.\frac{d\omega_{\star}}{d\tau}\right|_{\rm T} =\Gamma_{\rm T}t_\star/I_\star\Omega_\star$ is the dimensionless tidal torque, and $\chi=\Omega_\star^2 R_\star^3/G M_\star$ is a dimensionless ratio of stellar rotation speed to the surface Keplerian speed.
The dimensionless equation for the tidal torque is then 
\begin{equation}
    \gamma_{\rm T}(t)=- 3k \frac{t_{\star }}{t_{\rm conv}}\frac{q}{\beta_\star^2}\left(\omega_{\star}(t) - \left(\frac{r_{\star}}{\Tilde{a}(t)}\right)^{3/2}\left(\frac{M_\star+m}{\chi M_\star}\right)\right)\left(\frac{m}{M_\star}\right)^2  \left(\frac{r_{\star}}{\Tilde{a}(t)}\right)^6.
    \label{eq: tidal torque dimensionless}
\end{equation}
where we have used 
equation (\ref{eq: omega orb}) to eliminate $\Omega_{\rm orb}$ from equation (\ref{eq: tidal torque }) and rewritten it in terms of orbital separation $a(t)$.

\subsection{Combining magnetic braking and tides}\label{subsec:mb+tides}

Using a similar dimensionless prescription for the magnetic braking-induced wind torque as used for tides,  we have $   \gamma_{\rm W}=\left.\frac{d\omega_{\star}}{d\tau}\right|_{\rm W}=\Gamma_{\rm W}t_\star/I_\star\Omega_\star$, and the generalized equation for stellar spin evolution that combines both torques can then be written 
\begin{align}
    \frac{d{\omega}_{\star}(t)}{d\tau} &=     \gamma_{\rm W}(t)+\gamma_{\rm T}(t)(1-2{\rm H}(\Tilde{a}_{\rm crit}-\Tilde{a}(t)))\nonumber \\
    &= - {\omega}_{\star}(t) t_{\star}\frac{q}{\beta_\star^2}\frac{b_{r \star}(t)^2}{ m_{\star} \Tilde{u}_A(t)}\frac{B^2_{r,{\star}i}R^2_{\star}}{M_{{\star}i}u_{A,{\star}i}} \nonumber \\
    &- 3k \frac{t_{\star}}{t_{\rm conv}}\frac{q}{\beta_\star^2}\left(\omega_{\star}(t) - \left(\frac{r_{\star}}{\Tilde{a}(t)}\right)^{3/2}\left(\frac{M_\star+m}{\chi M_\star}\right)\right)\left(\frac{m}{M_\star}\right)^2  \left(\frac{r_{\star}}{\Tilde{a}(t)}\right)^6 \times \nonumber \\
    &  \times (1-2{\rm H}(\Tilde{a}_{\rm crit}-\Tilde{a}(t))).
	\label{eq: omegadot unified}
\end{align} 

To account for the demise of the companion when it reaches the critical orbital separation $\Tilde{a}_{\rm crit}$ defined as the larger of the Roche limit \citep{Eggleton1983} or stellar radius, we  include a multiplicative function $(1-2{\rm H}(\Tilde{a}_{\rm crit}-\Tilde{a}(t)))$ in the tidal torque term, where  ${\rm H}(\Tilde{a}_{\rm crit}-\Tilde{a}(t))
$ is the logistic function  given by
\begin{equation}
    {\rm H} (\Tilde{a}_{\rm crit}-\Tilde{a}(t))=\frac{1}{1+{\rm exp}[-0.2(\Tilde{a}_{\rm crit}-\Tilde{a}(t))]}.
\end{equation}
While the magnetic braking term 
always decreases the stellar spin, 
the influence of the tidal term 
depends on the orbital evolution of the secondary. If the secondary moves outward, then  
the
orbital angular momentum will increase and 
tidal force will act to spin down the star. But if the secondary moves inward, 
reducing the orbital angular momentum,
then tides 
increase the stellar spin. This evolution is
thought to be much more efficient for cool stars with outer convection zones than for hot stars with outer radiative layers \citep{Zahn2008}. 



\newcolumntype{L}[1]{>{\raggedright\arraybackslash}p{#1}}
\begingroup     
\everymath{\color{black}}  
\arrayrulecolor{black}   
\begin{table}[h]
  \centering
  \caption{Fixed model parameters}\label{tabparameters}
  \begin{tabular}{@{}l L{4.5cm} l l l @{}}
    \toprule
    & Parameter and Definition & Value & Method & References \\
    \midrule
    $\epsilon$ & X-ray saturation transition sharpness parameter  & [0.009–0.0036] & \parbox[t]{1.8cm}{ Theory 
    $\&$ Observation $\&$ Simulation }  & \parbox[t]{3.5cm}{\citep{Wright2011,Reiners2014,Wright2018}} \\
    $\lambda$ & exponent of the power-law dependence of magnetic starspot area
                   covering fraction on X-ray luminosity 
                 & 1/3 & Observation & \parbox[t]{3.5cm}{\citep{NicholsFleming2020}} \\
    $q$ & fraction of stellar mass anchored to the magnetic field & 1 & Theory &  Simple assumption \\
    $\beta_\star$ & radius of gyration & 0.458, 0.359 \footnotemark[1]  & Simulation & \citep{Claret2023} \\
    $k$ & apsidal motion constant & 0.155, 0.054 \footnotemark[1]  & Simulation & \citep{Claret2023} \\
    $\tau_c$ & convective turnover timescale & 
    \parbox[t]{4cm}{$\log \tau_c =2.33
- 1.50 \left(\frac{M}{M_\odot}\right)
+ 0.31 \left(\frac{M}{M_\odot}\right)^{2}$ }
    & \parbox[t]{1.8cm}{Observation $\&$ Simulation } & \citep{Wright2018} \\
    \bottomrule
  \end{tabular}
  \footnotetext[1]{ for $M_\star=0.2M_\odot$  \& $M_\star=0.6M_\odot$.}
\end{table}
\endgroup

\section{Results}\label{sec:results}
\subsection{Description of  systems studied}\label{sec:Description of  systems studied}

To study M dwarf-BD and K dwarf-Jupiter mass companion systems 
we numerically solve the generalized system of equations  (\ref{eq:lxbr}), (\ref{eq:bfi}), (\ref{eq:lxmdot2}), (\ref{eq:ua}), (\ref{eq: adot_tilda }) and (\ref{eq: omegadot unified}) derived in section \ref{sec:methods}, which describe the stellar spin evolution when spin-down due to magnetic braking is coupled to tides for regime II. To distinguish between fully and partially convective cases, we vary stellar mass and take different values for certain fixed model parameters, while solving the same set of equations. A number of  previous works indirectly infer  that there may be difference in dynamo mechanisms between fully and partially convective stars \citep{Vidotto2014,See2015,Folsom2018}. Here we assume that dynamo-related differences may be higher-order corrections to the present effects. Incorporating this into our calculations  is 
beyond the present scope. 
For all investigations below, we compare cases with both magnetic braking and tides to cases with just magnetic braking. For the latter,  we solve  equation (\ref{eq:omegadot wind}) 
instead  of equations (\ref{eq: adot_tilda }) and (\ref{eq: omegadot unified}).
Our model parameters  are given in Table \ref{tabparameters} 
and the initial
system properties for each case are given in Table \ref{tab1}. Using these  values,  
we vary the companion mass and fix the initial value of the orbital separation, or vice versa.  We define the initial orbital separation $a_{\rm i}$ as a fraction $p_{\rm f}$ of the corotation radius  $ a_{\rm i} =p_{\rm f} R_{\rm co}$, where  the corotation radius is given by $ R_{\rm co}=(G(M_\star+m)/\Omega^2_\star)^{1/3}$.

\begin{table}[b]
 \centering
\caption{Orbital system initial values 
}\label{tab1}%
\begin{tabular}{@{}llllll@{}}
\toprule
  & System 1  & System 2  & System 3\footnotemark[1] & System 4 & References \\
   &  &  & A, B  &  & \\
\midrule
$M_\star/M_\odot$  & 0.2 & 0.2   & 0.2  & 0.6  \\
$P_\star/\rm Days$ & 0.1 & 0.3  & 6  & 10  & \citep{popinchalk2021evaluating}\\
$B_{\rm p \star}/\rm G$   & 100 & 50  & 200  & 5  & \citep[e.g.,][]{Vidotto2014,See2015,Kochukhov2020}\\
$\Dot{M_\star}/\Dot{M}_{\odot}$ & 0.1  & 0.1   & 0.1 & 1  & \citep{Wood2021}\\
$m/M_{\rm J}$  & [15 - 75] & [15 - 75]  & [15 - 75], 75 & [2 - 16]
\\
$p_{\rm f} 
$   & 6 & 2.9   &  0.5, [0.5 - 0.95]  & 0.5
\\
\botrule
\end{tabular}
\footnotetext[1]{
For system 3 we consider two cases:  
for case A we change $m$ and fix $p_{\rm f}$ and for case B, we fix $m$ and change $p_{\rm f}$ while keeping all the other parameters and system properties unchanged. }
\end{table}


\subsection{Spin evolution for fully convective stars}\label{sec:spin evolution for fully convective stars}
\begin{center}
    \begin{figure}[!h]
        \centering
    	\includegraphics[scale=0.73]{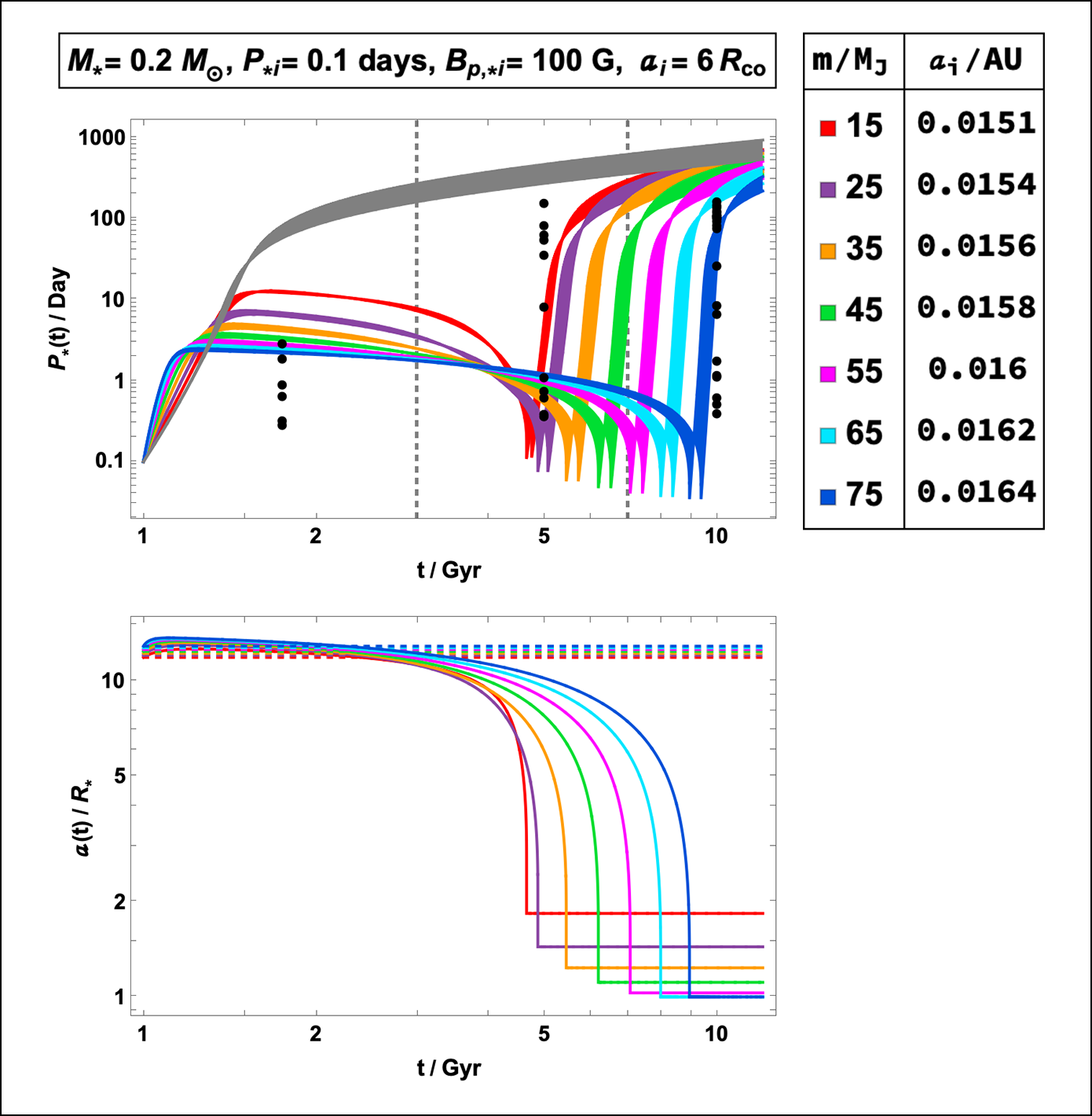}
    \caption{\textbf{Evolution of rotation and orbital separation 
    for System 1.} We use a $0.2M_\odot$ mass M-dwarf with $0.1$ days initial period, $100 \rm G$ magnetic field and companion BD at $a_{\rm i}=6 R_{\rm co}$ initial orbital separation. Different colors 
    represent varying BD companion masses between $15-75M_{\rm J}$. 
    Top panel:  spin evolution due only to magnetic braking (grey) and coupled to tides (colored).  Cusps represent engulfment points. The thickness of the grey and colored curves is determined by the highest and lowest $\epsilon$ values (see text). Black dots represent fully convective M dwarfs with measured rotation periods and estimated ages from Galactic kinematics \citep{irwin2011Angular}. The vertical grey dashed lines separate M-dwarfs in the thin, intermediate and thick disk populations with corresponding age ranges of $0.5-3$ Gyr, $3-7$ Gyr and $7-13$ Gyr.
    The x-axis and y-axis are normalized to 1 Gyr and a 1-day rotation period, respectively.
    Bottom panel: orbital separation for solutions without tides (dashed) and  coupled to tides (solid) analgous to the top panel. 
    The x-axis is normalized to 1 Gyr, and the y-axis is normalized to the stellar radius $R_\star$.}
    \label{fig:0.1}
    \end{figure}
\end{center}

The top and bottom panels of Fig. \ref{fig:0.1} show the time evolution of the rotation period and orbital separation for system 1 of Table \ref{tab1}.
For initial values we take a  rotation period of $P_\star =0.1$ days, a magnetic field  $B_{\rm p ,\star i}=100 \rm G$ and orbital separation  $a_{\rm i}=6 R_{\rm co}\approx 0.015\rm A
U$. We consider cases of companion BD masses between $15-75M_{\rm J}$, each represented by a different color. Mass accretion processes  are not considered in the current study, and may  be important for large companion masses.  We take $75M_{\rm J}$ BD dwarf as our largest companion mass case for visualization and to serve as a conceptual bridge towards future work involving stellar binaries and mass transfer.
Black dots are data points from \citep{irwin2011Angular} with measured rotation periods, and estimated ages from galactic kinematics for 41 field M-dwarfs. This data set  typifies the observed trend, and provides a simple comparison to our results, 
given the availability of both rotation periods and ages in the sample. In Fig. \ref{fig:0.1}, the vertical grey dashed lines represent boundaries between Galactic thin, intermediate, and thick disk M-dwarf populations with assigned ages $0.5-3$ Gyr, $3-7$ Gyr, and $7-13$ Gyr, respectively.  The data points in Fig. \ref{fig:0.1} use the  average value of these disk sub-population ages, since we do not have exact age measurements.
The grey area of the top panel corresponds to stellar spin-down without tidal interaction.
Solid and dashed lines in the bottom panel correspond to  cases with and without tidal interactions, respectively.
The thickness of the solution curves (grey and colored) depends on the highest and lowest $\epsilon$ values in equation (\ref{eq:lxbr}), which controls the sharpness of the transition in X-ray luminosity from the saturated (independent of stellar rotation) to the unsaturated (increases with  stellar rotation) regimes.

For Fig.\ref{fig:0.1}, the initial orbital separation exceeds the corotation radius.  All  companions therefore first move outward, gaining angular momentum from the star, and spinning it down faster 
than the case with only magnetic braking (grey curve). As the star spins down, the corotation radius increases ($R_{\rm co} \propto  \Omega_{\star}^{-2/3}$), eventually exceeding the orbital separation and causing the companion to migrate inward. 
Comparing  the two extreme companion mass cases exemplifies the relative evolution of  different mass cases. The heaviest companion (blue curve) extracts angular momentum much faster from the star than the lightest one (red curve), causing a faster transition from outward to inward migration. During the inward migration, the star gains angular momentum from the orbit until  the companion is fully tidally shredded, after which spin-down from magnetic braking dominates. 
 In the top panel of Fig. \ref{fig:0.1}, the dips \footnote{Double dips result from  solving cases with  different $\epsilon$ values that bound a range in $\epsilon$; sufficiently different $\epsilon$ values give discernibly different engulfment points.} correspond to the engulfments
 that occur when the companion reaches the  Roche limit or stellar radius, if the latter exceeds the former. In the bottom panel, the flat solid lines  correspond to  post engulfment evolution. 


We find, therefore, that a close BD companion can significantly delay the spin-down of less magnetically active M dwarfs  until tidal shredding. 
Such BD+M-dwarf systems can thus produce old M dwarf populations with short rotation periods ($ < 10$ days). In contrast, without a companion, the stellar wind removes enough angular momentum  via magnetic braking to increase the period to hundreds of days,  leading to  fast and slow rotator branches at fixed age. 
 Moreover, the rapid spin down after BD engulfment could explain the lack of observed M dwarfs  with intermediate rotation periods for the subpopulation of M dwarfs.
\begin{center}
    \begin{figure}[t!]
        \centering
    	\includegraphics[scale=0.73]{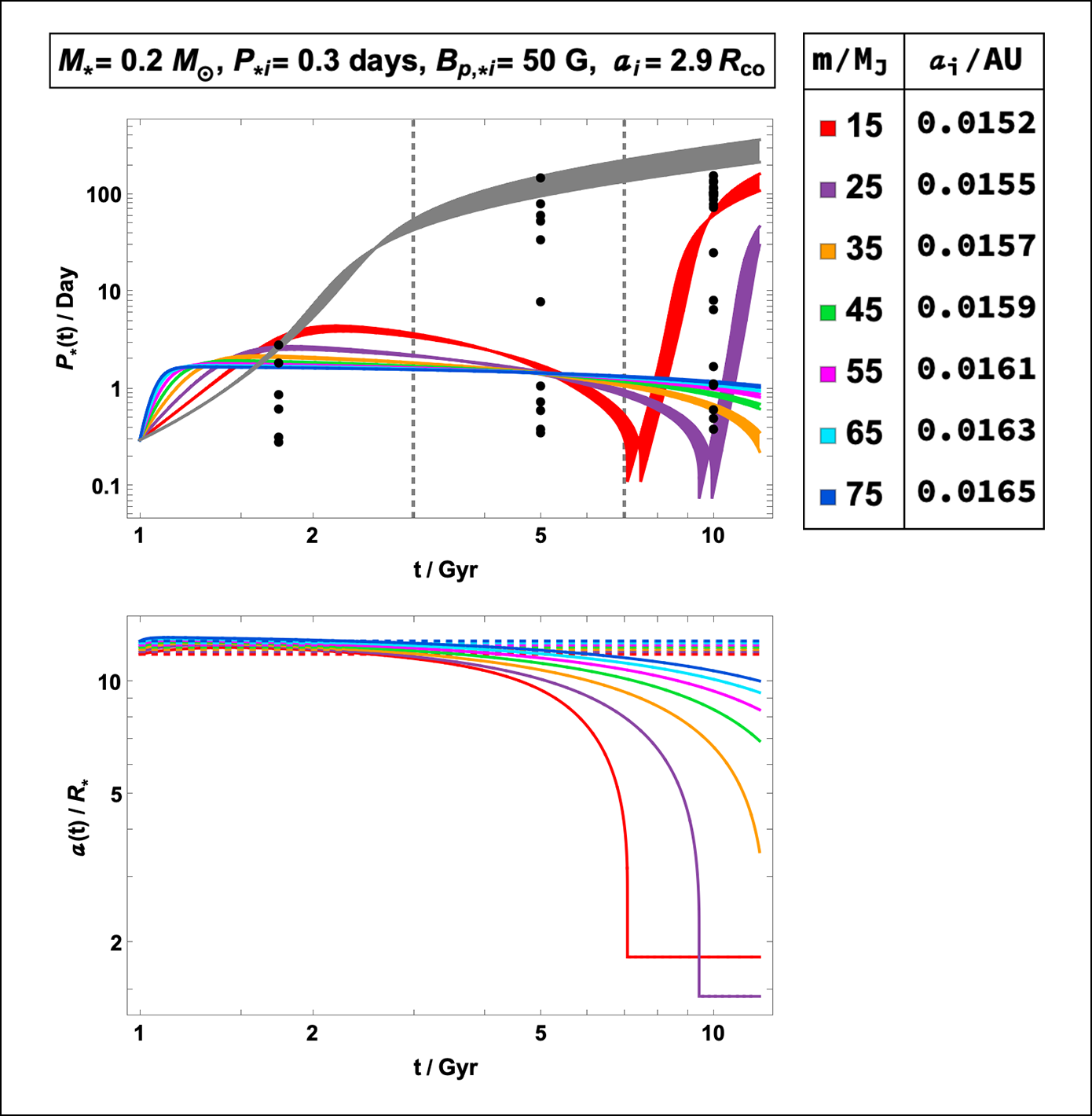}
        \caption{\textbf{Evolution of rotation 
    for System 2.}  We use an M-dwarf with the same stellar and  companion BD mass ranges as in Fig. \ref{fig:0.1}, but with  initial stellar rotation period of $0.3$ days, $50\rm G$ magnetic field, and  orbital separation at $a_{\rm i}=2.9 R_{\rm co}$. colors and axes are same as Fig. \ref{fig:0.1}.}
    \label{fig:0.3}
\end{figure}
\end{center}
In addition to companion mass and orbital separation, the spin evolution is  sensitive to the initial stellar rotation and magnetic field. Given that the combination of quantities needed for initial values in our model is not yet known for most  M dwarfs, we identify values that produce model predictions that capture the observed gap 
best, while remaining within the observationally constrained range for each stellar property.   
For system 2 in Fig. \ref{fig:0.3} compared to Fig.\ref{fig:0.1} (system 1), we change the initial rotation period to $0.3$ days and decrease the magnetic field  to $50 \rm G$. We keep the same companion mass range and start at $a_{\rm i}=2.9 R_{\rm co}\approx 0.015\rm A
U$ initial orbital separation. 
The spin evolves similarly to the previous case of Fig. \ref{fig:0.1}, but slower. As such, we would not observe tidal shredding of heavier companions within the age of the universe. For companions more massive than $30M_{\rm J}$, 
M-dwarf spin-down is delayed and the star
remains a fast rotator throughout its MS lifetime. To further illustrate the sensitivity of the spin evolution to the initial magnetic field strength and rotation period, we  present complementary cases to the Figs. \ref{fig:0.1} and Fig. \ref{fig:0.3} in the supplementary material, varying the initial magnetic field strength and rotation period separately (see Figs. \ref{fig:supl1} and \ref{fig:supl2}).

\begin{center}
    \begin{figure}[!t]
        \centering
    	\includegraphics[scale=0.73]{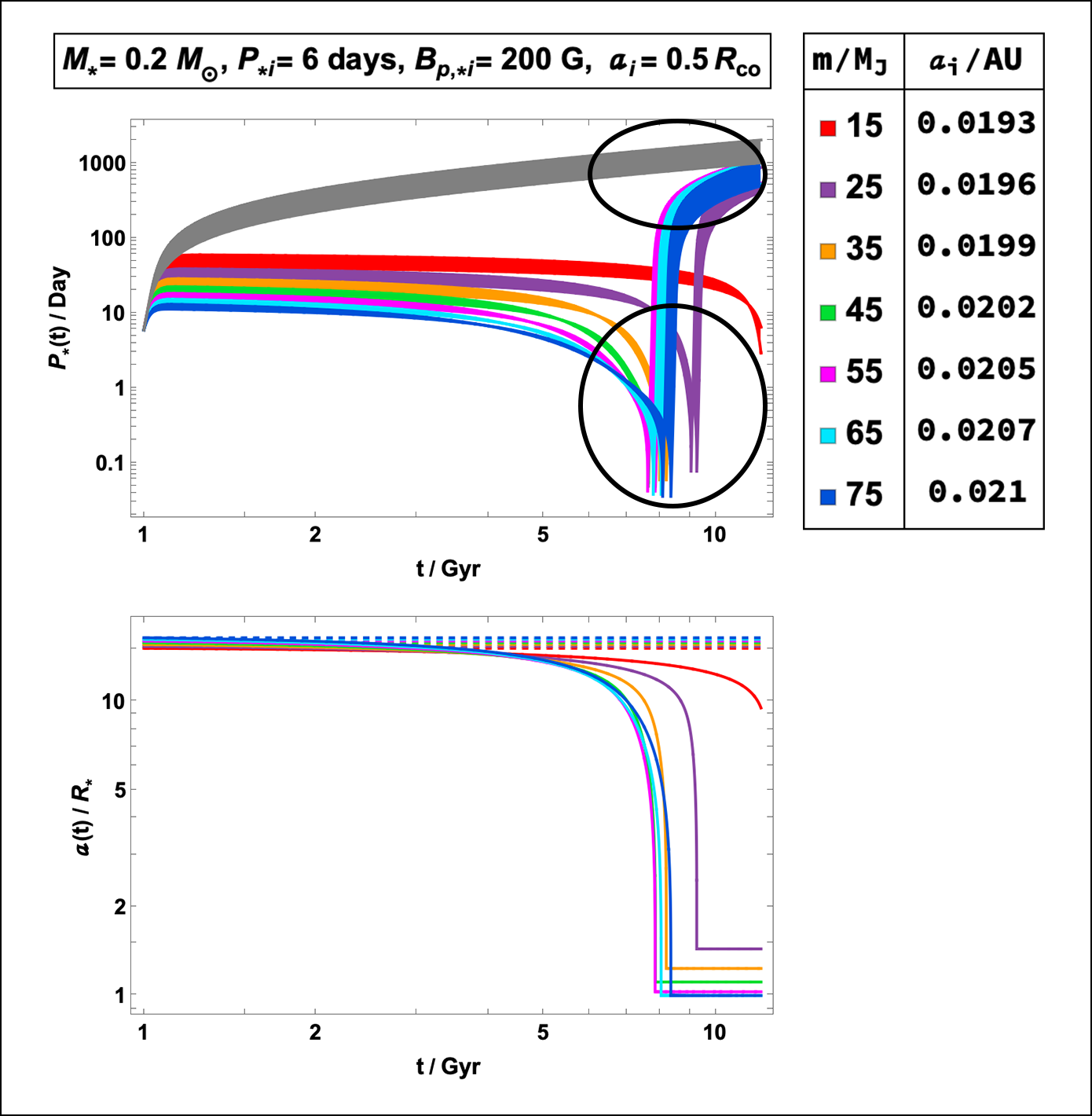}
    \caption{\textbf{Evolution of rotation and orbital separation 
    for System 3A.} 
   We use an M-dwarf with initial period of $6$ days, $200 \rm G$ magnetic field, and 
   orbital separation at $a_{\rm i}=0.5 R_{\rm co}$. 
   This causes  earlier inward migration of all the companions compared to Fig. \ref{fig:0.3} but also eventual engulfment.  Top panel: black circles  qualitatively illustrate the bimodal behavior of the M-dwarf rotation period during certain ages.  
   colors and axes are same as  Fig. \ref{fig:0.1}.}
    \label{fig:5}
\end{figure}
\end{center}

For systems 1 and 2, we focused on M-dwarf+BD dynamical evolution for  M dwarf initial rotation periods $P_\star < 1$day. In Fig. \ref{fig:5}, we show results for 
system 3A which starts with an M-dwarf 
of 
$P_\star =6$ days and $B_{\rm p, \star i}=200 \rm G$,  for the same companion BD mass range as in Fig. \ref{fig:0.3}
at the orbital separation of $ a_{\rm i}=0.5 R_{\rm co}$.  Similar to Fig. \ref{fig:0.1},  Fig. \ref{fig:5} shows that 
 the M dwarf initially spins down quickly from  magnetic braking, reaching a quasi-equilibrium state $|\gamma_{\rm W}(t)|\simeq|\gamma_{\rm T}(t)|$, where $\gamma_{\rm W}(t)$ and $\gamma_{\rm T}(t)$ are the dimensionless stellar wind and tidal torque, respectively (equation \ref{eq: omegadot unified}).  The companions migrate inward, spinning up the M dwarf but eventually  engulfing.
 The M dwarf subsequently returns to magnetically mediated spin-down. As in Fig. \ref{fig:0.1}, this can explain a rapid spin-down observed for M dwarfs \citep{Newton2016}. Here again, not many M-dwarfs would be expected  along the steep rising curves on this plot, leading to a gap and
 fast and slow rotator populations indicated by the two schematically circled regions.  Since initially tidal influence is not strong enough to significantly delay stellar spin-down or spin it up, the gap predicted from Fig. \ref{fig:5} would occur over a much shorter age range of only  $7-10$ Gyr once companions get closer to the star, compared to that predicted in Figs. \ref{fig:0.1} and \ref{fig:0.3}.


\begin{center}
    \begin{figure}[!t]
        \centering
    	\includegraphics[scale=0.73]{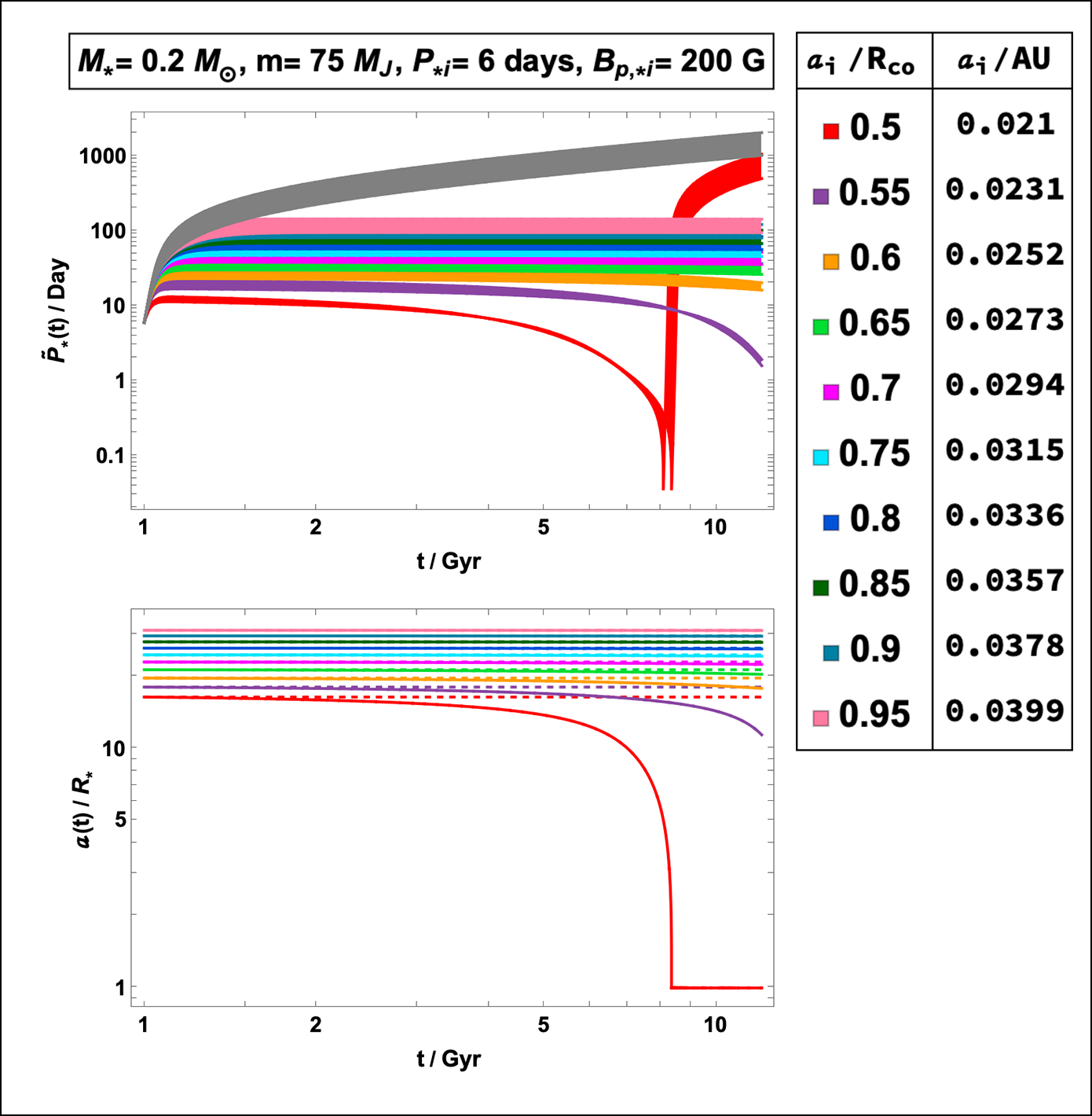}
    \caption{\textbf{Evolution of rotation and orbital separation 
    for System 3B.} We use an M-dwarf with the same stellar properties as in Fig. \ref{fig:5}  and  a $m=75M_{\rm J}$  companion BD. Here different colors represent  different initial orbital separations varying between $0.5-0.95 R_{\rm co}$. colors and axes are same as Fig. \ref{fig:0.1}.}
    \label{fig:52}
\end{figure}
\end{center}

\begin{center}
    \begin{figure}[!t]
        \centering
    	\includegraphics[scale=0.73]{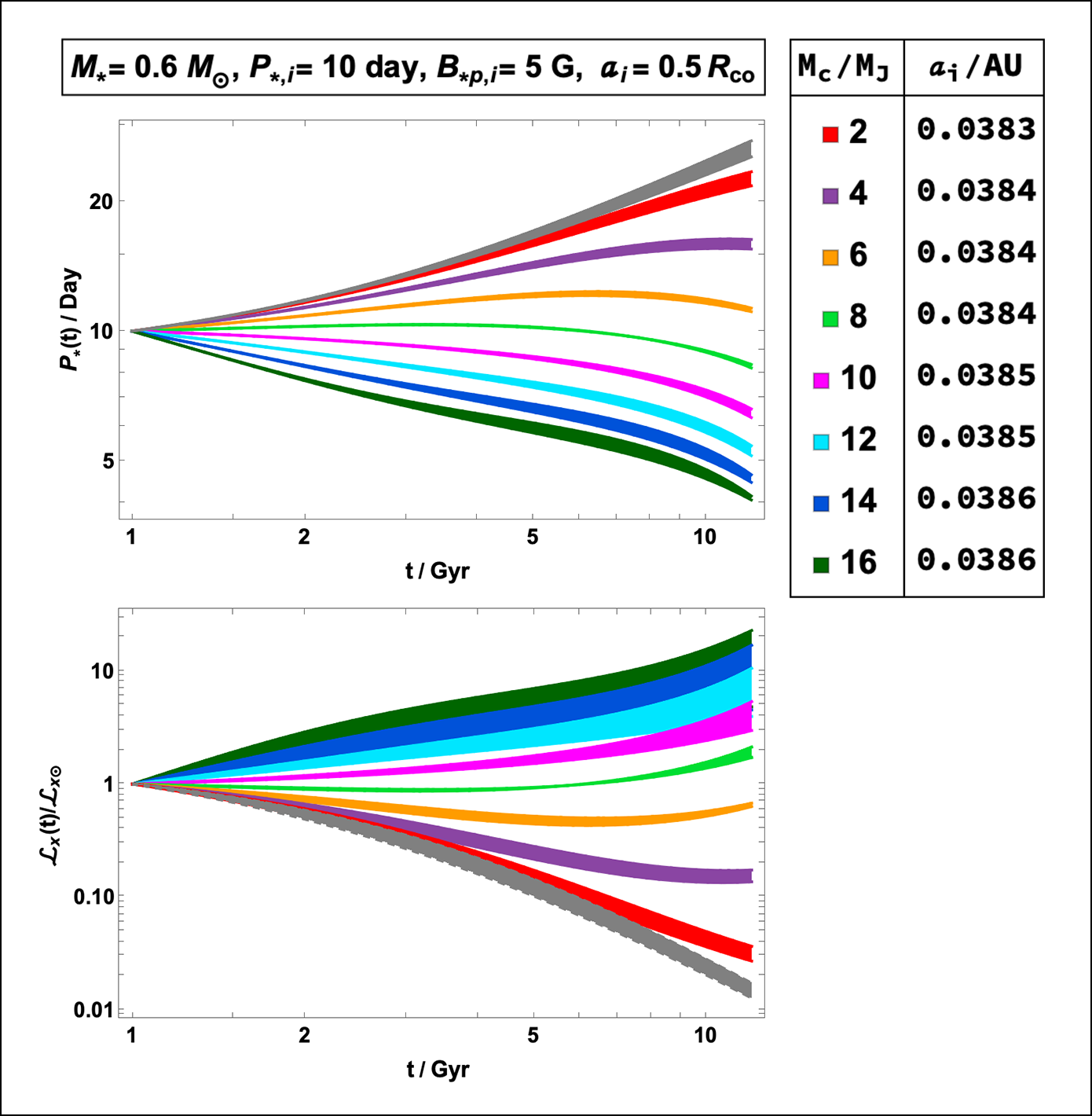}
    \caption{\textbf{Evolution of rotation period and X-ray luminosity $l_x$
    for System 4.} We use a $0.6M_\odot$ mass K-dwarf with a $10$ day initial period, $5 \rm G$ magnetic field   and  an initial orbital separation   $a_{\rm i}=0.5 R_{\rm co}$. Different colors in each panel represent different companion masses between $2-16M_{\rm J}$. Top panel: colors and axes are the same as  Fig. \ref{fig:0.1}. Bottom panel: The x-axis is normalized to 1 Gyr, and the y-axis is normalized to the solar X-ray luminosity $\mathcal{L}_{{\rm x}\odot}.$}
    \label{fig:MKnLum}
\end{figure}
\end{center}

For system 3B in Fig.\ref{fig:52} compared to system 3A of Fig.\ref{fig:5}, we take the companion mass to be a heavy BD of $75M_{\rm J}$ to maximize the influence of sub-stellar companion tides, and  vary the orbital separation from $a_{\rm i}=0.5 R_{\rm co}$ to $a_{\rm i}=0.95 R_{\rm co}$. 
Only when the companion starts at $a_{\rm i}\leq 0.5 R_{\rm co}$ it can spin up the star before engulfment. Even for heavy BD companions,  starting out farther than that cannot prevent the M dwarf from  spin-down. 
Wide enough orbital separations produce mutually similar solution curves. Thus a uniform distribution of orbital separations will not produce a uniform spacing of period-age evolution curves.  The particular empirical distribution of planet-star orbital separations will produce a distinct signature in the distribution of period-age evolution curves. In turn, a uniform distribution of  period-age evolution curves would  correspond to a  specific non-uniform planet-star orbital separation distribution. 

 In our model we use a simpler prescription for tidal interactions and do not take into account structural evolution of the host star or dynamical tides as compared to previous work \citep{Gallet2013,Gallet+Bouvier2015,Amard2016,Garraffo2018,benbakoura2019evolution,ahuir2021dynamical, Pezzotti2025}.
Although we have focused on lower mass star systems, our simpler self-contained analytic  approach, when applied to solar-like stars, captures similar  trends for the influence of tides. See for example Fig. 10 from \cite{ahuir2021dynamical} and our Fig. \ref{fig:0.1}, but note the difference in system parameters. The ultimate limitations of accuracy from  different assumptions and  ingredients that go into these different frameworks  makes the present simpler approach complementarily useful.
A more detailed comparison between these different approaches across  parameter regimes is beyond the present scope.


\subsection{Spin evolution for partially convective stars}\label{sec:Spin evolution for partially convective stars}
Jupiter mass companion planets are more common for K-dwarfs than for M-dwarfs.  Here we consider system 4 of Table \ref{tab1}, 
a $0.6M_\odot$ mass star with a $10$ day rotation period, $5\rm G$ magnetic field, and a companion mass range $2M_{\rm J}\leq m\leq 16M_{\rm J}$ at an initial orbital separation $a_i=0.5R_{\rm co}$. Fig. \ref{fig:MKnLum} shows the resulting time evolution of the rotation period and  X-ray activity. The plots exemplify that 
high-mass close companions can spin up a host star (top panel) and thereby increase X-ray activity (bottom panel). Our results agree with previous studies (e.g \cite{ilic2022tidal, Pezzotti2025}), which show that massive close-in planets can  increase X-ray activity and shorten rotation periods.
Compared to the M-Dwarf+BD systems of Fig. \ref{fig:0.3}, this system produces more evenly distributed curves for  a uniform companion mass distribution,  so the gap discussed for those cases is not predicted here.







\section{Challenges in identifying companions and associated observational implications} 
\label{sec:Discussion}

\subsection{Substellar and Stellar companions of M Dwarfs}
In general, observations of M dwarfs and their companions face several challenges due to their faintness,  distances, and stellar variability. Combined with observational limitations, sample sizes and selection biases \citep{Obermeier2016}, these factors are likely to result in the underestimation of M dwarfs with companions \citep{Cifuentes2025carmenes}. Currently, the occurrence of known brown dwarfs at small separations ($\leq 1\rm {AU} $) around low-mass M dwarfs is approximately $1.3\%$ \citep{Winters+2019} and is even less for Jupiter mass planets \citep{Endl2006}. However, the stellar  binary multiplicity rate for M dwarfs ranges from $28\%$ to $16\%$,  depending on M dwarf mass 
\citep{Winters+2019}. Recent observations \citep{Cifuentes2025carmenes} suggest additional binary candidates which, if confirmed, could raise the multiplicity rate to $40.3\%$. This potentially  significant increase suggests that more high-resolution observations are needed. In this study,  we focus on M dwarfs with substellar companions for simplicity and to highlight the significance of tidal interactions in the context of spin evolution. It is a first step toward the future generalization of our model to stellar binary systems and the inclusion of mass accretion processes, given the high multiplicity rate of M dwarfs.
Also,  considering  M dwarf systems with multiple small planetary companions, which are more frequent, could collectively produce enough tidal torque to significantly influence or even delay M dwarf spin-down.

\subsection{Companion survival }\label{sec:Discussion}

Disk instability models suggest that Jupiter-mass planets can form rapidly around M dwarfs \citep{Mercer2020A}, so that the absence of Hot Jupiters around M dwarfs could be due to detection biases or post formation processes. High-energy radiation from active M dwarfs may lead to companion photoevaporation \citep{Kurokawa2014}, while in less active M dwarf systems, such companions can migrate inward and eventually get engulfed by the host star \citep[e.g.][]{Siess1999,Nordhaus2013}. In this study, we focus on less active M dwarfs, but according to our crude estimation based on energy-limited formalism from \citep{Erkaev2007, Lopez2013}, even for highly active M dwarfs, companions with the masses studied here are unlikely to  evaporate within 10 Gyr. Therefore,   the planetary engulfment scenario seems a more probable outcome for our systems. Increased activity level and rotation of the star can be used as a proxy for tidal effects \citep{Ilic2022}, since 
inward migration and engulfment can lead to significant increase in stellar rotation   \citep{Siess1999, Carlberg2010,Oetjens2020} and  activity \citep{Siess1999,Kashyap2008, Ilic2023}. 
Depending on stellar and companion properties, there may also be a temporary change in the surface abundance of lithium \citep{Carlberg2010}. \citep{Cabezon2023} shows that BD engulfment by a solar-like star can cause lithium enhancement by a factor of 20-30, depending on the BD mass. Thus, in addition to stellar rotation and activity observations, high-precision abundance surveys are needed for companion host stars as indirect evidence of the recent substellar companion engulfment.

\subsection{Challenges for gyrochronology 
}

Gyrochronology can in principle  
determine isolated low-mass MS stellar ages with reasonable precision \citep{barnes2003rotational}. But
our study shows how the one-to-one mapping on which this method depends is broken with companions.
Moreover, although we have explored scenarios with just one  companion, 
 the occurrence rate of planets around M dwarfs is at least 3.0 planets per star, potentially exacerbating the influence of tides on spin \citep{Tuomi2019} and age-rotation degeneracy.
Recalibrating gyrochronology  to account for tides would be desirable to extend its utility beyond isolated stars. 
This is challenging but maybe possible because,
for a large range of companion BD masses over a large age range, the period evolves little  (see Fig. \ref{fig:0.3} ). Also,  
 the curves for heavier companions are  close together, and less sensitive to the influence of a specific companion mass.

\section{Conclusions and Outlook}\label{sec:Conclusion}

We have studied the combined influence of tides
and magnetic spin-down of M-Dwarfs  for  different companion BD masses, initial orbital separations, stellar rotations and magnetic field strengths.  We find that plausible M-Dwarf+BD binary interactions offer a possible explanation  for the observed period distribution
gap.
Fig. \ref{fig:0.1}, Fig. \ref{fig:0.3} and Fig. \ref{fig:5} reveal this gap differently.
In Fig. \ref{fig:0.1}, the spin-evolution tracks exhibit a period gap, followed by sharp spin-up spikes during tidal shredding, followed by steeply rising curves that result from the rapid M-dwarf spin-down after companion engulfment. 
 Fig. \ref{fig:0.3} shows that the spin evolution tracks are separated by a period gap over their full age for most of the companion masses. In Fig. \ref{fig:5} the 
 M-dwarfs initially spin  down, but the companions move inward fast enough to  rapidly spin-up the star, consistent with  a period gap (E.g. \cite{Newton2016}).  After companions are engulfed,  the stars rapidly spin down by magnetic braking.  Since the  spin-down transition  is fast,  the fraction of stars therein is small and the overall period distribution becomes approximately  bimodal with fast and slow rotator populations.


For a $0.6M_\odot$ partially convective star however,  Fig. \ref{fig:MKnLum}  shows that  
a uniform distribution of companion masses produces more evenly spaced curves than the more pronounced slow and fast rotator population branches 
for  fully convective M dwarf cases of Figs. \ref{fig:0.1}, \ref{fig:0.3}  and  \ref{fig:5}. This is independent of any magnetic topology changes  or changes in core to envelope magnetic coupling, which have elsewhere been proposed to influence spin evolution \citep{Garraffo2018,David2022, Cao2023}.
In our framework, the presence
of 
a pronounced gap between slow and fast rotators for partially convective stars compared to fully convective M-dwarfs would reveal differences in companion mass and orbital separation distributions between the two stellar population classes, and may constrain the statistics of stellar-planetary system formation outcomes. 

Observations show that lower mass M dwarfs   have a  companion to primary mass ratio $m / M_\star$, that increases exponentially with decreasing mass for M-dwarfs of masses  $M_\star \lesssim 0.25 M_\odot$. In addition,  massive M-dwarfs of $M_\star \gtrsim 0.35 M_\odot$
lack companions at very close separations \citep{Winters+2019}. 
These trends would imply statistically larger tidal torques and a
larger gap between slow and fast rotators for fully convective M-dwarfs (see equation (\ref{eq: tidal torque1})), and 
   relative to their partially convective counterparts.  

Knowing either companion mass or orbital separation 
 from observations for a specific system can eliminate some degeneracy between the two in our model.  The predicted spin evolution is  more sensitive to orbital separation than to companion mass and so  observations that  pin down the  orbital separation distributions will help to constrain the predicted fraction of M-dwarfs whose spin is tidally influenced. For example, a uniform distribution of companion separations could imply that in $\sim 80\%$ of systems  the companion would not influence spin substantially (Fig. \ref{fig:52}) whilst a higher fraction of companions at smaller separations would exacerbate the relative population of old M-dwarfs with anomalously short periods. 
Complementarily, the observed gap for old M-dwarfs can be used to constrain BD or planet mass-orbital separation distributions around fully convective M dwarfs.  If tides cause the gap, this would signature the presence of BD companions and prioritize particular systems  for companion searches in follow-up observations.
We have included only equilibrium tides in this work. Dynamical tides can enhance the tidal influence and  offset the spindown of even more magnetically active M-dwarfs than we have considered. This would be particularly important for modeling specific M dwarf-BD systems  were data  to become available.






Finally, although observations show no sharp weakening of the rotation-activity at the fully convective M dwarf boundary (around M3 spectral type) they 
  show a sharp drop in activity  for stars later than spectral type  M8  \citep{Mohanty+Basri2003Apj,Mohanty+2002ApJ}. 
 \citep{Mohanty+2002ApJ} suggested that magnetic braking is  less effective in stars later than M8 because a low ionization fraction leads to a high enough magnetic diffusivity such that the magnetic field 
 decouples from the plasma near the surface. 
 This implies a correlation between low activity and short spin periods in these  M-dwarfs that is also consistent with our calculations, as tides do little to exacerbate spin down for these cases. We would then predict no significant population of long period low mass M-dwarfs later than $M8$, unless magnetic braking can   survive  in stars with these  low ionization layers.

\section*{Acknowledgments}



KK acknowledges support from a Horton Graduate Fellowship from the Laboratory for Laser Energetics.  We acknowledge support from 
National Science Foundation grant 
PHY-2020249 and DOE grant DE-SC0021990.
EB acknowledges the Isaac Newton Institute for Mathematical Sciences, Cambridge, for support and hospitality during the programme "Frontiers in dynamo theory: from the Earth to the stars", supported by EPSRC grant no EP/R014604/1. This material is based upon work supported by the Department of Energy [National Nuclear Security Administration] University of Rochester “National Inertial Confinement Fusion Program” under Award Number(s) DE-NA0004144.

\bibliographystyle{apsrev4-1}

\bibliography{spindown}
\section*{Supplementary information}
\label{sup}


\begin{center}
    \begin{figure}[!h]
        \centering
	\includegraphics[scale=0.75]{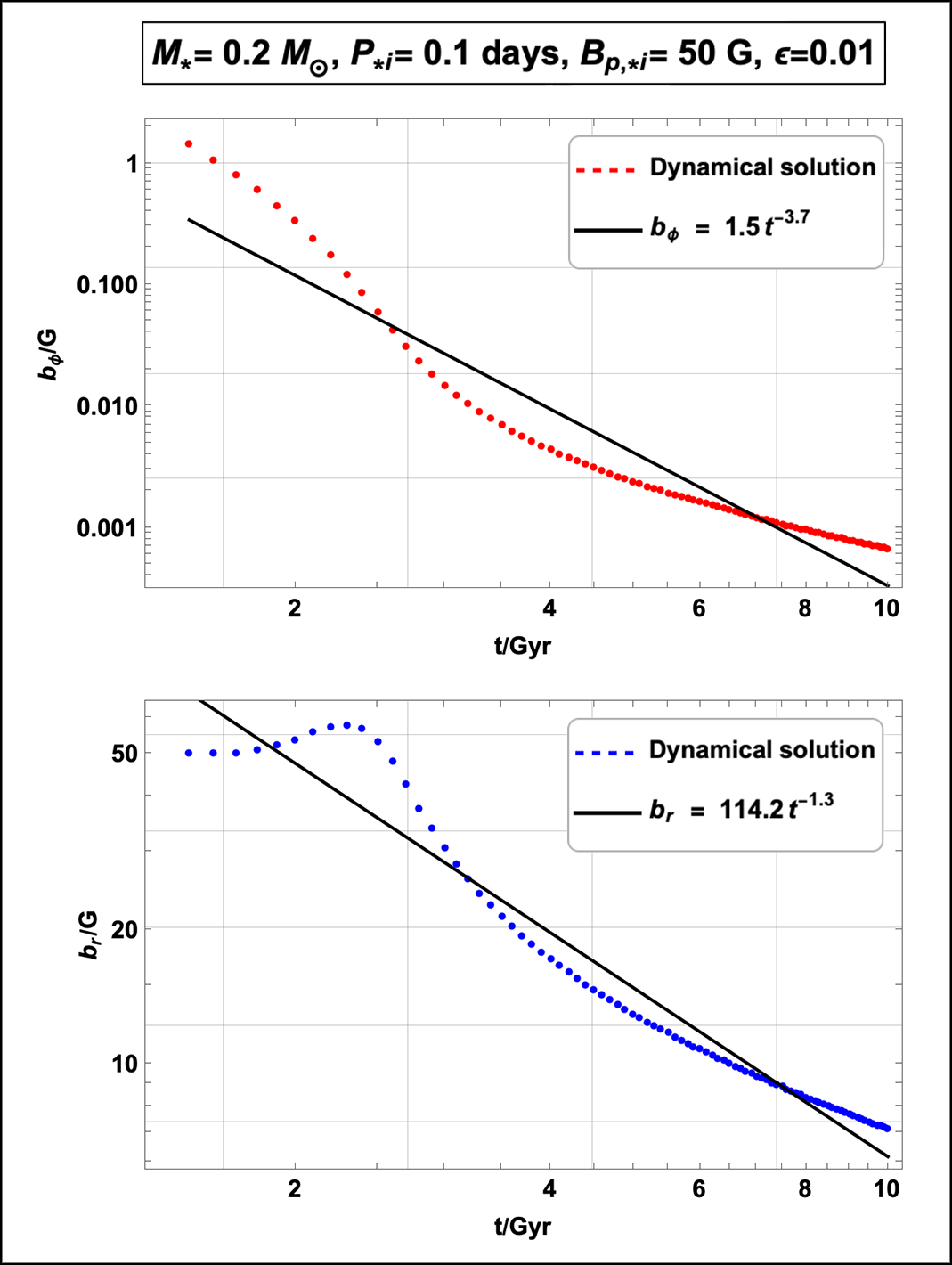}
    \caption{{\color{black}Evolution of toroidal  and radial magnetic fields for solving our dynamical equations in the top and bottom panels in red and blue respectively for $\epsilon=0.01$. For the initial stellar  values, we used a mass of $M=0.2M_\star$, a rotation period  $P_\star =0.1$ days, and a magnetic field  $B_{\rm p,\star i}=50 \rm G$.  At each point, on these curves, the tangent  lines would represent a power low scalings  of the form $ b_\phi(t) \propto t^{\xi_1(t)}$ and $ b_r(t) \propto t^{\xi_2(t)}$. The black lines represent approximations of  single best fit line that approximates   $\xi_1(t),\xi_2(t)$ from $t\approx
 1.5-10$ Gyr.  From these line scalings shown,  we solve for $b_\phi(t)$   as a function of $b_r(t)$  and further account for $|B_r(t)| \propto b_r(t)^{3}$ to infer $\eta=\xi_1/(3\xi_2)$ for relation between the toroidal field and predicted observed
radial magnetic field $b_\phi(t)\propto C |B_r(t)|^\eta $. In this case $\eta=0.96.$. This rough scaling is reasonably consistent with observed ranges \citep{See2015}}.}
    \label{fig:suplfit}
\end{figure}
\end{center}

\begin{center}
    \begin{figure}[!h]
        \centering
	\includegraphics[scale=0.73]{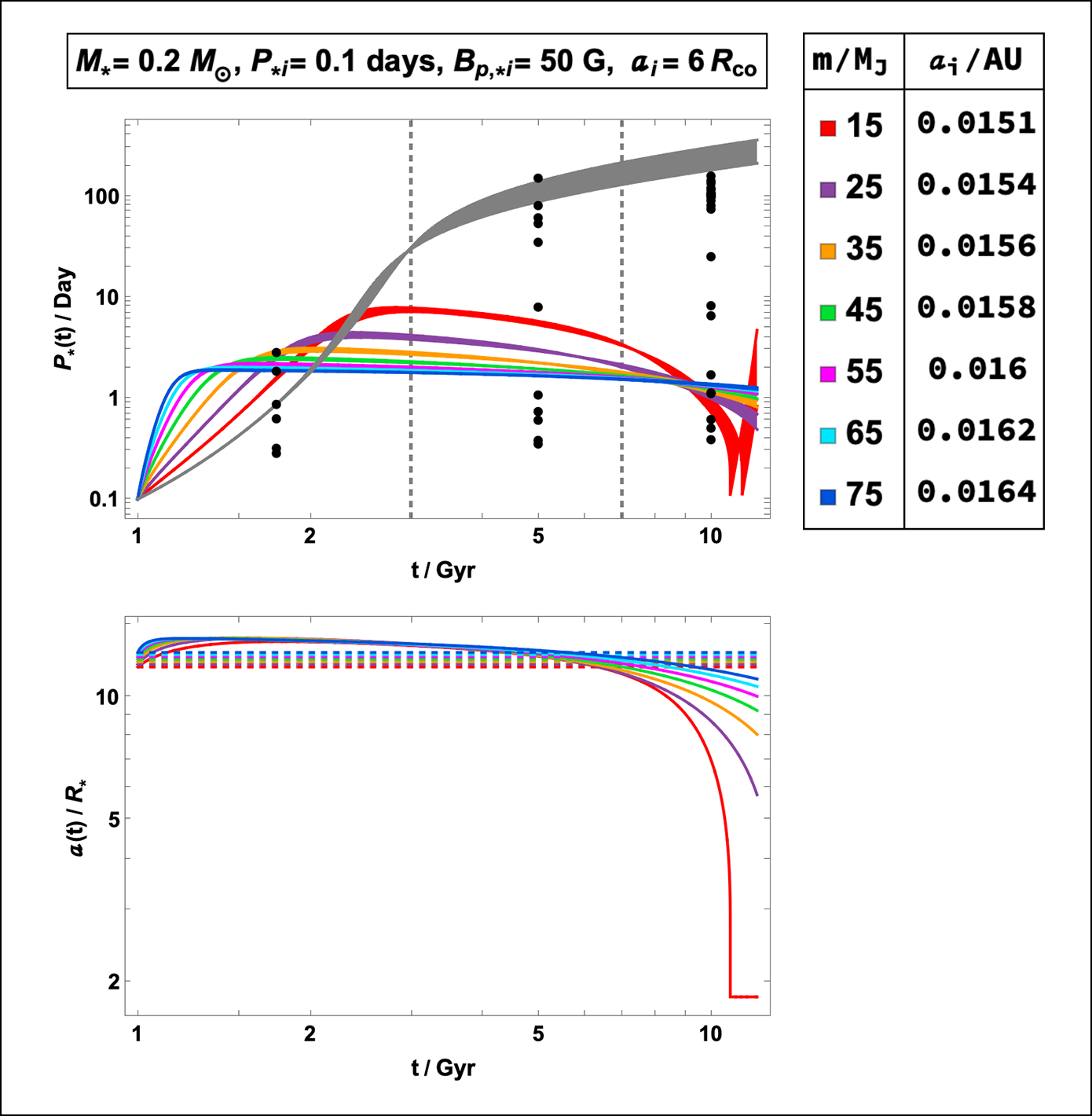}
    \caption{ M-dwarf with the same initial properties and companion BDs as in Fig. 1, but with an initial $50\rm G$ magnetic field. Colors and axes are the same as Fig. 1.}
    \label{fig:supl1}
\end{figure}
\end{center}

\begin{center}
    \begin{figure}[!t]
        \centering
	\includegraphics[scale=0.73]{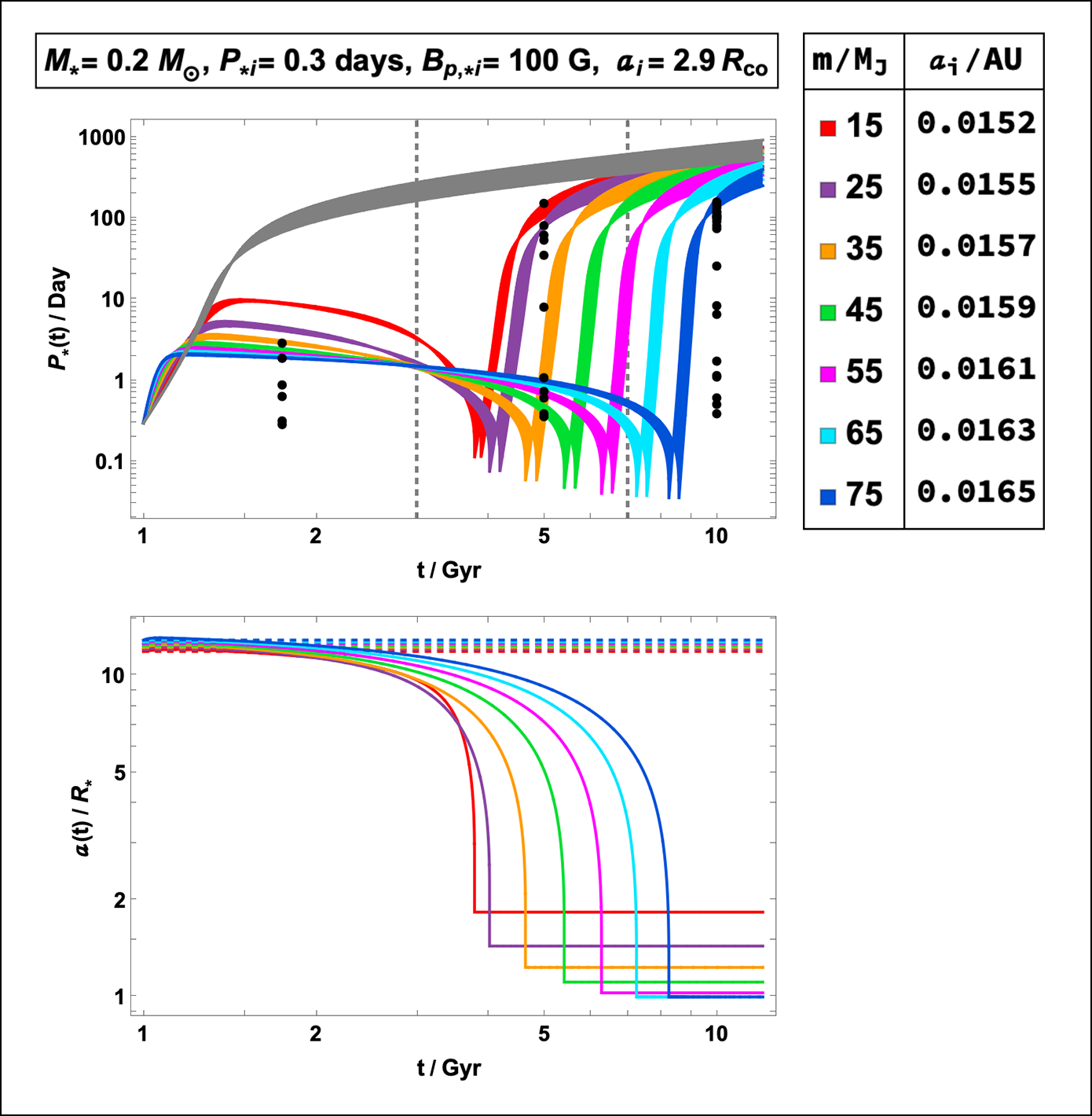}
    \caption{ M-dwarf with the same properties and companion BDs as in Fig. 2, but with initial $100\rm G$ magnetic field strength. Everything else is illustrated and normalized similarly to Fig. 2.}
    \label{fig:supl2}
\end{figure}
\end{center}

\end{document}